\documentclass{naturep}

\usepackage{amssymb}
\usepackage{amsmath}
\usepackage[mathscr]{eucal}
\usepackage{mathalpha}
\usepackage{mathtools}
\usepackage{eucal}
\usepackage{lipsum}
\usepackage{amsfonts}
\usepackage{graphicx}
\usepackage{algorithmicx}
\usepackage{algorithm}
\usepackage{subcaption}
\usepackage[noend]{algpseudocode}
\usepackage{bbm}
\usepackage{hyperref}
\usepackage{lineno}
\usepackage[margin=0.6in]{geometry}
\usepackage{ragged2e}
\usepackage{setspace}
\usepackage{longtable}
\usepackage{caption}
\usepackage{hyperref}
\usepackage{caption}
\usepackage{anyfontsize}
\usepackage{setspace}
\usepackage{float}
\usepackage{booktabs}
\usepackage{multirow}
\usepackage{xcolor}
\usepackage{tablefootnote}
\usepackage{url}

\usepackage{newfloat}
\usepackage{hyperref}
\usepackage{cleveref}

\DeclareFloatingEnvironment[name={Extended Data Figure},listname={Extended Data Figure}]{extfigure}
\Crefname{extfigure}{Extended Data Figure}{Extended Data Figures}

\DeclareFloatingEnvironment[name={Extended Data Table},listname={Extended Data Table},]{exttable}
\Crefname{exttable}{Extended Data Table}{Extended Data Tables}

\makeatletter
\let\saved@includegraphics\includegraphics
\AtBeginDocument{\let\includegraphics\saved@includegraphics}

\makeatother

\graphicspath{{Figures/}}

\makeatletter
\def\thickhline{%
  \noalign{\ifnum0=`}\fi\hrule \@height \thickarrayrulewidth \futurelet
   \reserved@a\@xthickhline}
\def\@xthickhline{\ifx\reserved@a\thickhline
               \vskip\doublerulesep
               \vskip-\thickarrayrulewidth
             \fi
      \ifnum0=`{\fi}}
\makeatother
\newlength{\thickarrayrulewidth}
\title{\begin{flushleft}{\begin{spacing}{1}
Deep Learning-based Frozen Section to FFPE Translation
\end{spacing}}\end{flushleft}}

\begin{document}

\maketitle
\begin{spacing}{1}
\vspace{-15mm}
\noindent Kutsev Bengisu Ozyoruk$^{1}$, Sermet Can$^{1, 2}$, Guliz Irem Gokceler$^{1}$,  Kayhan Başak$^{6}$, Derya Demir$^{7}$, Gurdeniz Serin$^{7}$, Uguray Payam Hacisalihoglu$^{8}$, Emirhan Kurtuluş$^{1}$, Berkan Darbaz$^{1}$,  Ming Y. Lu$^{3,4,5}$, Tiffany Y. Chen$^{3,4}$, Drew F. K. Williamson$^{3,4}$, Funda Yılmaz$^{7}$, Faisal Mahmood$^{*,3,4,5}$ and Mehmet Turan$^{*,1}$
\begin{affiliations}
  \item Institute of Biomedical Engineering, Bogazici University, Istanbul, Turkey
  \item UK Biocentre, Tilbrook, England
 \item Department of Pathology, Brigham and Women's Hospital, Harvard Medical School, Boston, MA
 \item Cancer Data Science Program, Dana-Farber Cancer Institute, Boston, MA
 \item Cancer Program, Broad Institute of Harvard and MIT, Cambradge, MA
\item Sağlık Bilimleri University, Kartal Dr.Lütfi Kırdar City Hospital, Department of Pathology, Istanbul, Turkey
\item Faculty of Medicine, Department of Pathology, Ege University, Izmir, Turkey
 \item Istanbul Yeni Yuzyil University Medical Faculty, Gaziosmanpasa Hospital, Pathology Department

 \end{affiliations}

\noindent\textbf{Code / Package:} {\color{blue}\href{https://github.com/DeepMIALab/AI-FFPE}{https://github.com/DeepMIALab/AI-FFPE}}\\

\end{spacing}
\begin{spacing}{1}
\noindent\textbf{*Correspondence:}\\ 
\noindent Mehmet Turan\\
Institute of Biomedical Engineering,\\ 
Bogazici University, Istanbul, Turkey\\
mehmet.turan@boun.edu.tr

\noindent Faisal Mahmood \\
60 Fenwood Road, Hale Building for Transformative Medicine\\
Brigham and Women's Hospital, Harvard Medical School\\
Boston, MA 02445\\
faisalmahmood@bwh.harvard.edu

\end{spacing}

\newpage

\noindent\textbf{\large{Abstract}}
\begin{spacing}{1}
\noindent Frozen sectioning (FS) is the preparation method of choice for microscopic evaluation of tissues during surgical operations. The high speed of procedure allows pathologists to rapidly assess the key microscopic features, such as tumor margins and malignant status to guide surgical decision-making and minimise disruptions to the course of the operation. However, FS is prone to introducing many misleading artificial structures (histological artefacts), such as nuclear ice crystals, compression, and cutting artefacts, hindering timely and accurate diagnostic judgement of the pathologist. Additional training and prolonged experience is often required to make highly effective and time critical diagnosis on frozen sections. On the other hand, the gold standard tissue preparation technique of formalin-fixation and paraffin-embedding (FFPE) provides significantly superior image quality, but is a very time-consuming process (12-48 hours), making it unsuitable for intra-operative use. In this paper, we propose an artificial intelligence (AI) method that improves FS image quality by computationally transforming frozen-sectioned whole-slide images (FS-WSIs) into whole-slide FFPE-style images in minutes. AI-FFPE rectifies FS artefacts with the guidance of an attention-mechanism that puts a particular emphasis on artefacts while utilising a self-regularization mechanism established between FS input image and synthesized FFPE-style image that preserves clinically relevant features. As a result, AI-FFPE method successfully generates FFPE-style images without significantly extending tissue processing time and consequently improves diagnostic accuracy. We demonstrate the efficacy of AI-FFPE on lung and brain frozen sections using a variety of different qualitative and quantitative metrics including visual Turing tests from 20 board certified pathologists. Additionally, we demonstrate that deep models trained on FFPE tissue can adapt to frozen sections which a much higher efficacy when using AI-FFPE.


\noindent 

\end{spacing}

\newpage
\begin{spacing}{1} 
\vspace{-3mm}

\begin{figure*}[!b]
\includegraphics[width=1\textwidth]{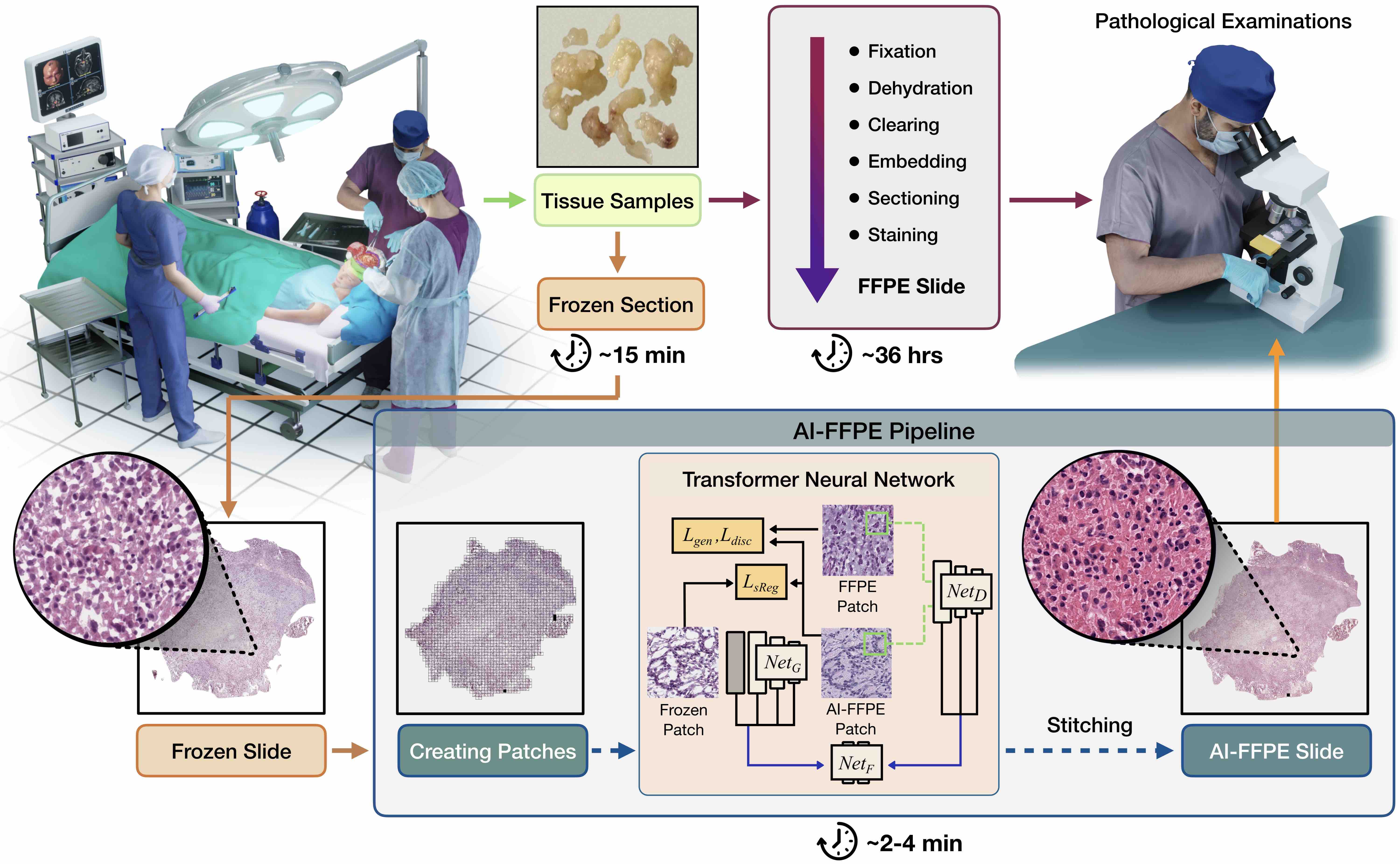}
\caption{\textbf{Workflow Overview.} Diagram summarising how AI-FFPE method fits into the routine preparation of surgically excised specimens for histopathological evaluation. The tissue samples are processed with either FFPE or frozen sectioning(FS). While the former takes around 36 hours, provides higher quality images and definitive diagnosis; the latter takes only around 15 minutes, is crucial for the intra-operative decision-making but results in significantly compromised image quality and diagnostic accuracy. Our rapid 2-4 minutes AI-FFPE tool is integrated into FS process as a last step, converting FS images into FFPE-style images to overcome FS-specific obstacles. AI-FFPE pipeline receives digitized high-resolution whole-slide FS histology images (FS-WSIs) as an input. For each FS-WSI, NxN mini-patches are created and the transformer neural network works on each patch separately to fix the chemical and mechanical artefacts. At the last step, patches are stitched back for the examination of the pathologists providing improved-quality FFPE-style images at a speed that is compatible with fast-pace of intra-operative histopathological examinations. 
} 
\label{fig:overview}
\end{figure*}

\noindent 
Histologic examination of tissue by a pathologist is the gold standard for the diagnosis of many diseases. Though this examination is most often performed on formalin-fixed paraffin embedded (FFPE) tissues for final diagnosis, a faster alternative called “frozen sectioning” is a crucial tool in the arsenal of surgeons and pathologists for intra-operative guidance of resections, usually for assessment of tumor margins, differentiation of malignant vs. benign lesions and intra-operative staging. The process of FFPE can take 12 to 48 hours, far exceeding the time-limits of routine intra-operative decision-making. Instead, pathologists use frozen sectioning (FS), immediate freezing and cutting of tissues, accelerating the process of preparing slides from hours to minutes. The trade off of this increase in speed, however, is the introduction of artefacts from freezing and cutting specimens, artefacts that can cause significant rates of discordance between frozen section and more accurate subsequent FFPE diagnoses \cite{Patil2015,Humberto2015,Rogers1987,Xiang2020}.

\noindent Frozen-sectioning artefacts are numerous and include distortion of cellular details and loss of tissue entirely due to ice crystal formation, folding and tearing of the delicate sections, as well as large variances in staining due to varying section thickness \cite{Brown2009,Jaafar2006} and contaminants in the staining reagents \cite{Chatterjee2014,Taqi2018}, see \textbf{\Cref{fig:artefacts}}. These mostly irreversible artefacts can severely distort the appearance of the tissue as compared to FFPE \cite{Ensel2015,PICHAT201873,Renne2019CryoembedderAP}, disguising malignant cells and making benign cells look atypical. Indeed, there are some tissues for which frozen section for intra-operative consultation is relatively uncommon, such as fatty tissues, due to these artefacts making reliable evaluations nearly impossible. However, FS is widely applied to the lung and brain tissues for rapid identification of unexpected masses, to differentiate malignant vs. benign and primary vs. metastatic lesions. Nevertheless, frozen-sectioned lung and brain cancer samples represent a significant diagnostic challenge to the histopathologists as documented by relatively high discordance rate between diagnostic results of frozen-sectioned and FFPE tumor samples' examinations. Resolving FS-FFPE disagreements, more precisely improving accuracy of diagnosis using FS samples offers significant impact on healthcare because lung cancers are the leading cause of cancer deaths and the decisions taken during the brain cancer surgeries could have tremendous consequences on patients' post-operative functionality and quality of life. Lung and brain tissues' and tumors' many distinct characteristics seem to cause above-mentioned increased discrepancy in diagnosis. Lung tissue, for example, harbors air sacs at large volumes which upon frozen-sectioning, particularly with thin sectioning, can collapse and lead to misdiagnoses of atelectasis and in some cases can result in the earliest form of lung cancer, \textit{in situ} adenocarcinomas, going unnoticed.  In brain, for instance, differentiation of less aggressive (low grade) subtype of the most common brain cancers, gliomas, from a benign condition called gliosis might be impeded by FS-introduced artefacts. Thus, relatively high FS-FFPE diagnostic discordance combined with the prospect of significant impact on human health makes lung and brains tumors' FS images prime candidates for improvement by AI. Furthermore, the diagnostic discordance rate data in the literature originate from academic medical centres with highly experienced and/or specialised experts on site. These rates are expected to increase in centres with smaller volumes and more generalist pathology teams because only a small proportion of samples examined in pathology departments are frozen-sectioned, limiting pathologists' experience of FS specimen examinations.  Thus, translation of FS images to FFPE-style images would provide pathologists with images that they have significantly more familiarity with. In summary, we addressed in this study the issues arising from high frequency of artefacts in FS tissue slides and limitations on gaining experience of FS samples by generating FFPE-like images from FS images with an ultimate aim of increasing the diagnostic accuracy of FS examinations to the levels of FFPE examinations. To achieve this goal, we designed a method, AI-FFPE, that utilizes deep style transfer approach based on generative adversarial networks (GANs).          
\begin{figure*}
	\centering
	\includegraphics[width=0.96\columnwidth]{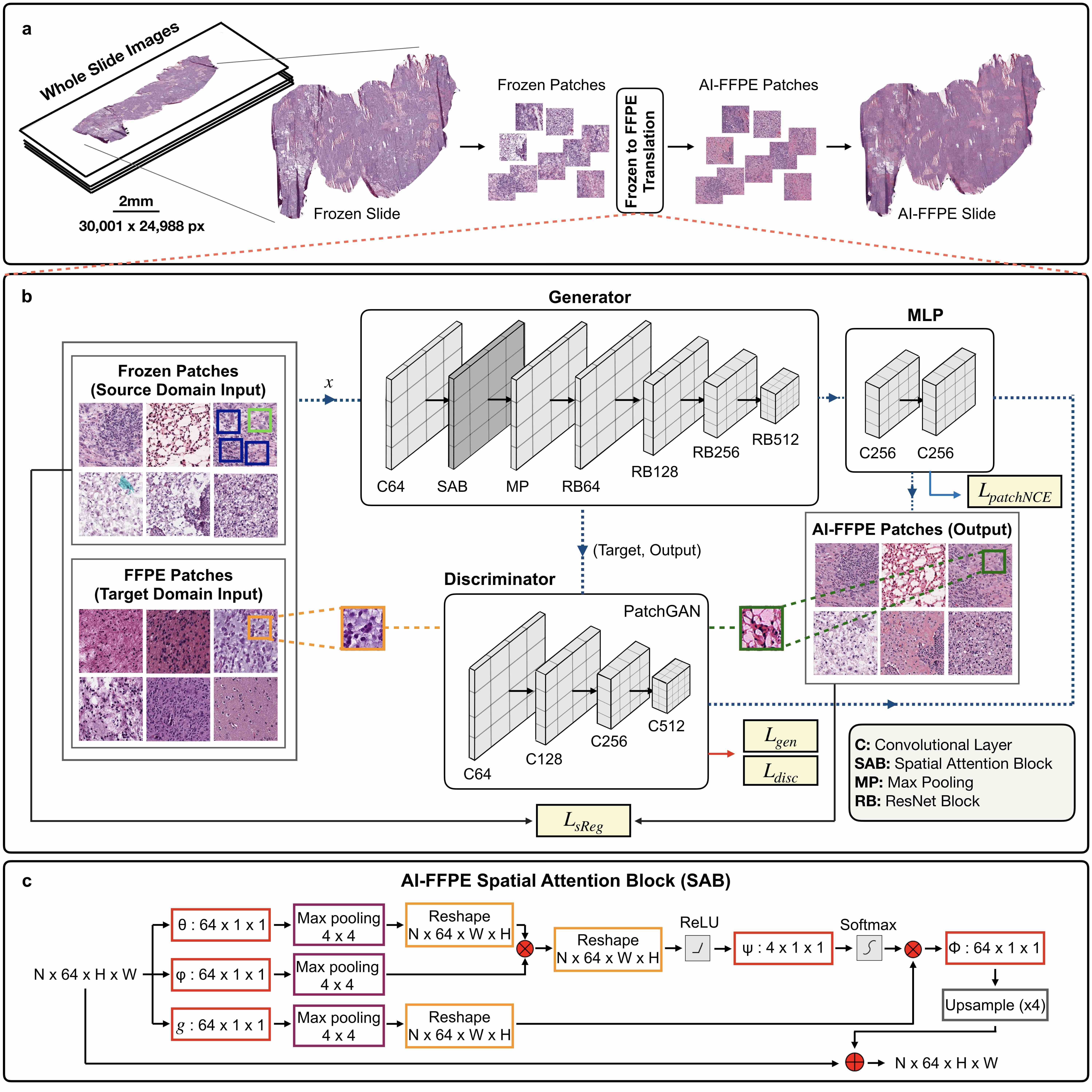}
	\caption{\textbf{AI-FFPE Method Architecture.} \textbf{a} Process overview. FS-WSIs are first cropped into 512x512 square patches for AI-FFPE networks to convert them into FFPE-like images which were, subsequently, stitched together to construct final WSIs to be examined by pathologists. \textbf{b} A more detailed diagram of AI-FFPE transformer networks. The input images from the source domain (FS) are fed into the generator whose output is then served to the discriminator after being concatenated with the image from the target domain (FFPE). To promote the output images' resemblance to the target domain, we use adversarial loss which is a weighted summation of $\mathcal{L}_{gen}$ and $\mathcal{L}_{disc}$. For the encouragement of content preservation by concentrating networks on commonalities between two domains, self regularization and patch-wise noise contrastive estimation loss are integrated into the overall objective function. A two-layer MLP network is added alongside the generator to improve the association between input and generated data. \textbf{c} The flow diagram of spatial attention block (SAB) embedded in ResNet-9 generator for the synthesis of the in-silico alternative of FFPE sections.  
	}
	\vspace{-4mm}
	\label{fig:summary}
\end{figure*}

\section*{\large{Results}}
\label{sec:results}

\vspace{-2mm}
\section*{ Optimisation of the AI Model for Frozen Section to FFPE Translation}

We constructed a novel unpaired neural-style transfer framework that is trained under the supervision of both patch-wise and pixel-wise signals where the content domain is FS and the style domain is FFPE images. We adopted an architecture typically found in GANs which consists of a ResNet-based generator and a PatchGAN discriminator with the Least Square GAN loss-of-contrastive learning. To design the final form of the algorithm, we assessed how integration of self-regularization (SR) constraint and spatial attention block (SAB), and the combination thereof effects the performance of the algorithm \textbf{\Cref{fig:ablation}}. Integration of SR-loss function enhanced nucleo-cytoplasmic contrast and staining quality, and also prevented unnecessary introduction of red blood cells into the image, a modification that could give false impression of bleeding into the tissue (\textbf{\Cref{fig:ablation} a, \& b}). SR-loss alone enabled the AI-FFPE to fill out the blank regions, however, this new function was executed in an relatively indiscriminate manner (\textbf{\Cref{fig:ablation} c \& d}). This inadequate selectivity of the algorithm has been resolved with the integration of an SAB modality. Interestingly, combination of SR-loss with SAB resulted not only in cumulative improvement but also in synergistic effect of incorporating textural details to the ECM. Rarely, combination of these modalities reduced the contrast between the nuclei and the cytoplasm. However, copious benefits of combining the two modalities grossly exceed its minimal disadvantages, therefore, we integrated SR-loss and SAB into the final version of the model. Detailed examples of how these two modalities improve the image features are shown in \textbf{\Cref{fig:ablation}} and a detailed illustration of the final AI-FPPE network architecture can be found in \textbf{\Cref{fig:summary}}.


\begin{figure*}[!b]
,
\includegraphics[width=\textwidth]{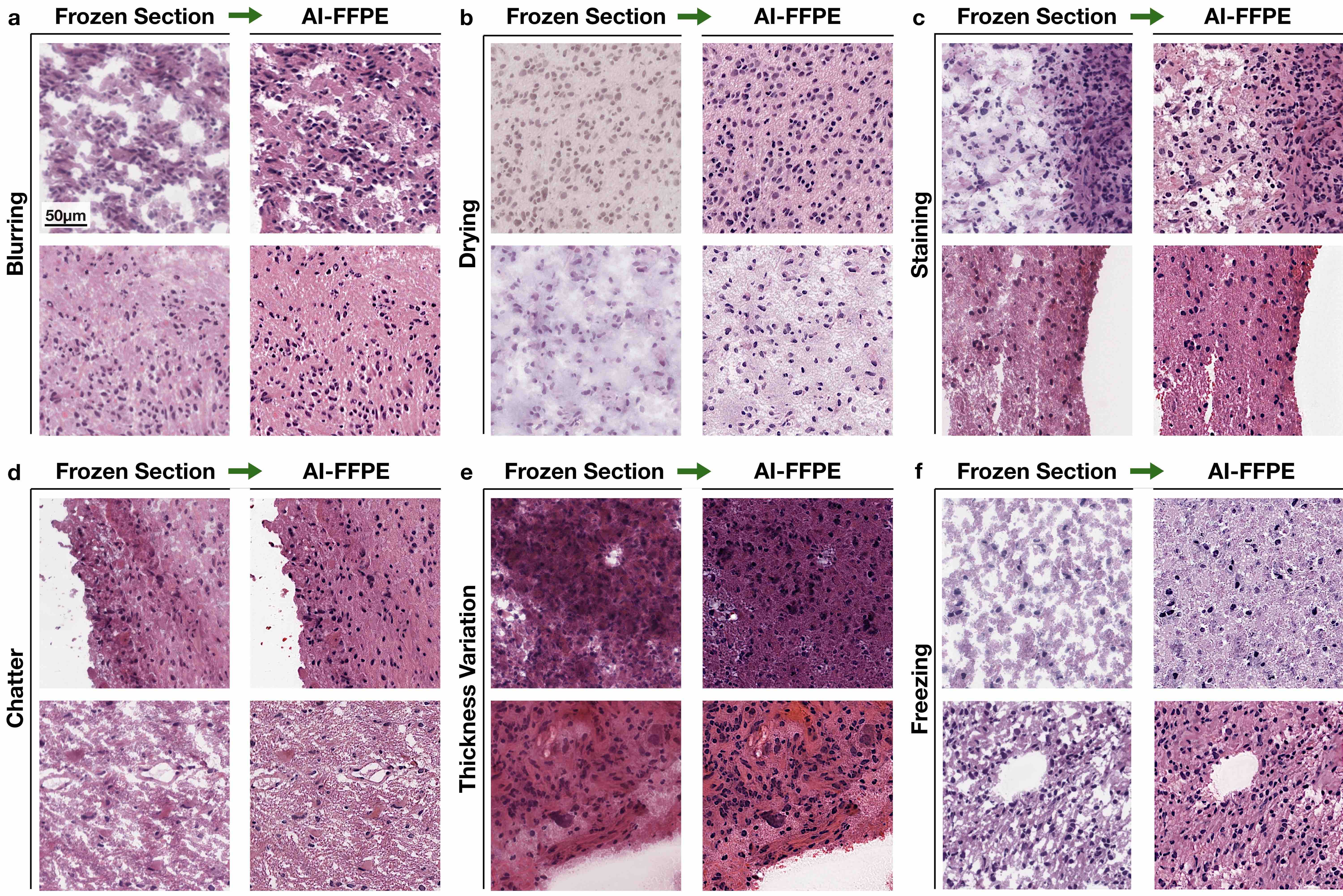}
	\caption{\textbf{Improvement of artefacts in the brain frozen sections.} Examples of AI-FFPE correction of different frozen-sectioning artefacts. The patches selected to represent one class of artefact, however these artefacts does not exist in isolation and different artefacts show overlapping features. AI-FFPE corrects these artefacts simultaneously.  \textbf{a} Reversal of blurring artefact with rectification of the losses in the ECM. \textbf{b} Distorted cellular details due to drying artefacts are corrected in the AI-FFPE column. Improvements in the staining quality and ECM texture are also present in the same column. \textbf{c} Correction of staining artefacts by increasing colour intensity, range and contrast, accompanied with minor improvements in the ECM. \textbf{d, e} Correction of chatter and thickness variation artefacts by enhancing nuclear borders in the thicker parts and the repair of ECM gaps in the thinner parts of the tissue. \textbf{f} Blank areas appear due to large ice crystal formations are restored. AI-FFPE also sharpened nuclear borders and improved staining quality.}
	\vspace{-4mm}
\label{Fig:BrainResults}
\end{figure*}


\section*{Evaluation of Model Performance}

\begin{figure*}[!b]
\includegraphics[width=\textwidth]{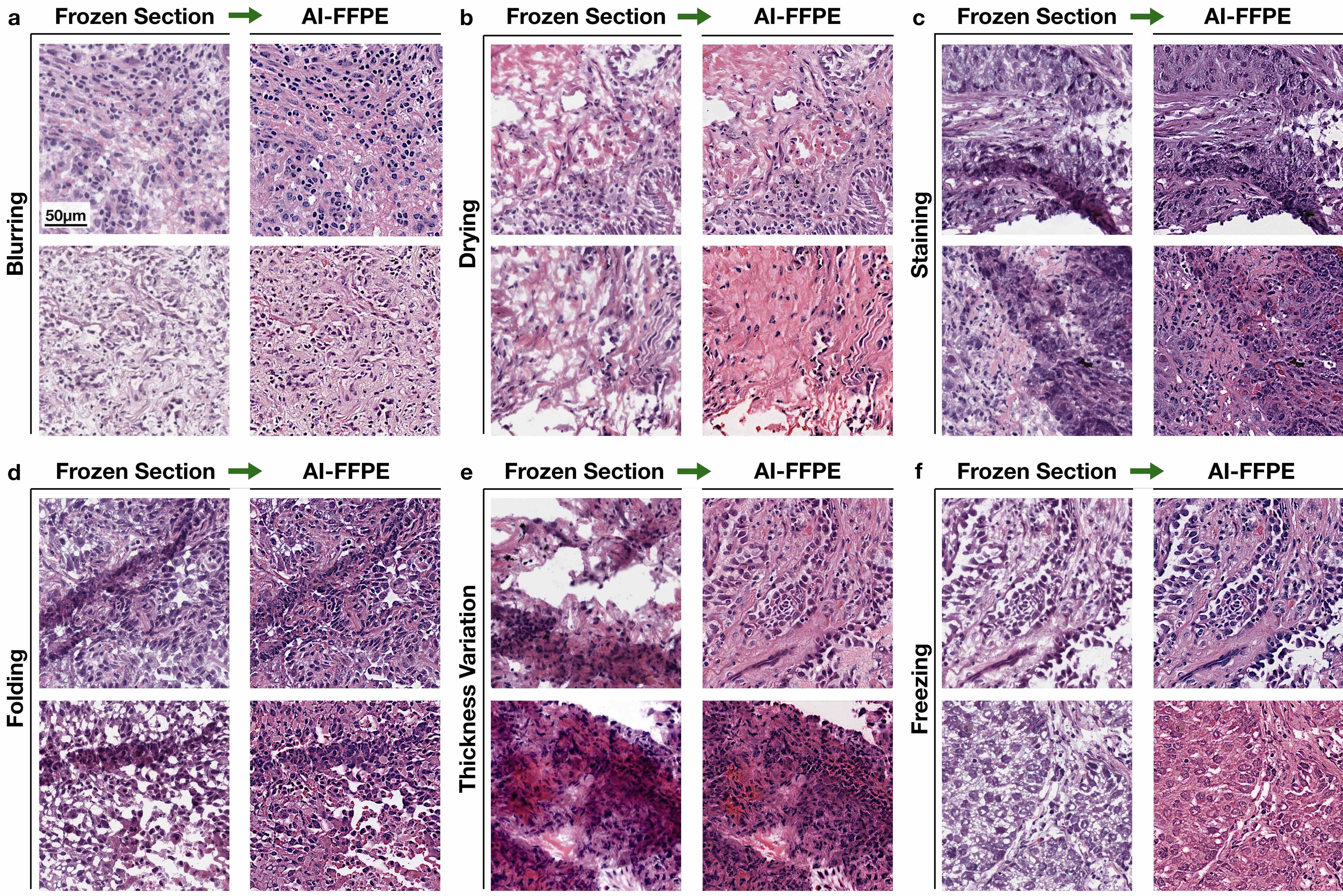}
	\caption{\textbf{Improvement of artefacts in the lung frozen sections.}  Similar to the brain images, artefacts highlighted in each patch coexist with other types of artefacts. \textbf{a} AI-FFPE enhanced the resolution of blurred patches. Other improvements, such as ECM repair and improved staining quality are also present. \textbf{b} Compromised cellular details due to drying artefacts and distorted ECM are restored together. \textbf{c} Improved colouring of the images with folding and thickness variation artefacts. AI-FFPE also significantly rectified image quality issues occurred due to the latter two. \textbf{d, e} Augmentation of the appearance of cells in the thicker areas. Upper row patches of "e" also show enhanced visualisation of tissue infiltrating lymphocytes \textbf{f} Empty spaces due to ice crystal formation in the ECM (both patches) and inside the cells (lower patch) are corrected and staining quality is improved.} 
	\vspace{-4mm}
\label{Fig:LungResult}
\end{figure*}

Frozen sectioning, when compared to FFPE, introduces additional and unique misleading artificial structures (histological artefacts) to the tissue (see \textbf{\Cref{fig:artefacts}}), a major reason. Therefore, we first examined if our model reverses these artefacts. Our results demonstrate that AI-FFPE efficiently corrects various frozen section artefacts in brain (\textbf{\Cref{Fig:BrainResults}}) and lung (\textbf{\Cref{Fig:LungResult}}) sections, such as freezing, cutting, drying and staining artefacts. These artefacts often exist concurrently and exhibit shared elements. AI-FFPE rectifies each type of artefact by resolving various image quality issues together. For instance, by increasing the prominence of nuclear borders and generating more pronounced ECM texture, our method showed outstanding efficiency in overcoming blurring artefacts in lung and brain tissue slides (\textbf{\Cref{Fig:BrainResults} a},  \textbf{\Cref{Fig:LungResult} a}). Similarly, in WSIs where regional cell densities have altered due to chattering (\textbf{\Cref{Fig:BrainResults} d}), folding (\textbf{\Cref{Fig:LungResult} d}) and thickness variation (\textbf{\Cref{Fig:BrainResults} e}, \textbf{\Cref{Fig:LungResult} e}) artefacts, AI-FFPE, by increasing the contrast between nucleus and cytoplasm, allows enhanced visualisation of individual cells in the thicker regions of the slides where cells appear in high density because of being stacked on top of each other.  Moreover, AI-FFPE reduces the differences in tissue thickness by filling the artificially occurring blank areas due to the freezing process. One of the most important benefits that AI-FFPE offers is the recovery of staining quality by improving the colour intensity, contrast and spectrum (\textbf{\Cref{Fig:BrainResults} c}, \textbf{\Cref{Fig:LungResult} c}). Also, frozen sectioning tend to produce drying artefacts which hinders the pathologists' ability to discern cellular structures, such as nuclear architecture and cytoplasmic borders, all of which are reversed by our method by restoring structural contrasts and improving colouring quality. AI-FFPE also corrects artefacts that are exclusive to frozen sectioning such as presence of blank areas due to the formation of ice crystals (\textbf{\Cref{Fig:BrainResults} f}, \textbf{\Cref{Fig:LungResult} f}). It is very important to note that while reversing such diverse types of artifacts individually or in combination, our method does not seem to introduce any misleading structures (\textbf{\Cref{Fig:BrainResults}, \textbf{\Cref{Fig:LungResult}}}).  

We also compared the artefact-correcting performance of our method (final version, w/o SR-loss, w/o SAB) with generic image translation models of FastCUT and CycleGAN on the brain (\textbf{\Cref{fig:all_models_brain}}) and the lung (\textbf{\Cref{fig:all_models_lung}}) WSIs. The results show that our model, custom-designed for FS to FFPE translation, clearly outperforms these generic image-translation models, and the modalities that we have added to our method substantially contributed to its artefact-correcting performance.


In order to quantify AI-FFPE model's efficiency in transformation of FS images to FFPE-style images, we employed Frechet inception distance (FID), a well established metric of similarity for assessment of AI generated images' proximity to the target domain images. We compared AI-FFPE with generic image translation models of CycleGAN, CUT and FastCUT using the dataset that is detailed in the "Dataset Description, Online Methods" sub-section. We found that AI-FFPE's FID values were the lowest, hence images generated by AI-FFPE model were the closest to the real FFPE images, both for brain (AI-FFPE: 29.81 vs CUT: 32.28, FastCUT: 34.42, CycleGAN: 69.43) and lung (AI-FFPE: 28.15 vs  CUT: 35.49, FastCUT: 35.71, CycleGAN: 39.19) images, demonstrating the superiority of our model in translation of the images from FS to FFPE domain.

\begin{figure*}[!h]
	\centering
	\includegraphics[width=\columnwidth]{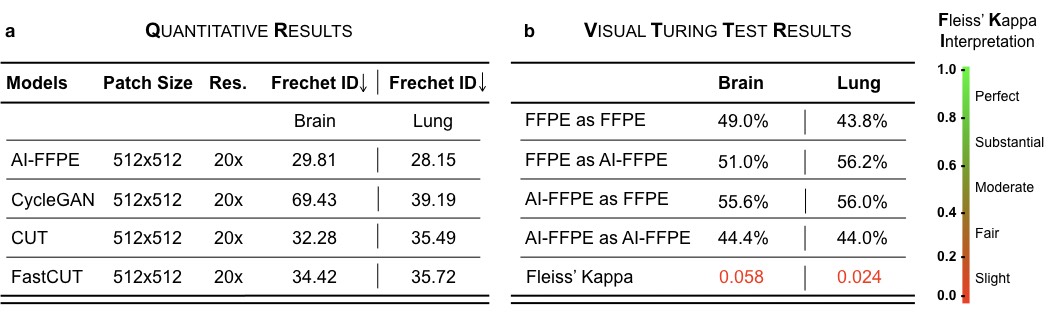}
	\caption{\textbf{Assessment of the AI-FFPE generated images' similarity to the real FFPE images.}   \textbf{a}    Comparative FID results on 38,165 FS lung patches from 187 patients and 25,721 FS brain patches from 141 patients. AI-FFPE is compared with generic image translation models: CycleGAN, CUT and FastCUT. Lower FID scores for AI-FFPE, compared to the other models, indicate the  generation of images with higher resemblance to the FFPE domain on both lung and brain samples.  \textbf{b} {Results of Visual Turing Tests involving 20 pathologists evaluating a collection of 512 x 512 sized 25 AI-FFPE and 25 FFPE patches from brain and lung tissues. The pathologists were asked to determine whether the images are real FFPEs or FFPE-like images generated from FS images by AI-FFPE model. The results show that the pathologist could not distinguish between FFPE and AI-FFPE images further corroborating AI-FFPE's efficiency in transforming FS images to FFPE format. The Fleiss' kappa values of 0.024 for lung and 0.058 for brain tissues indicate a lower "slight" inter-observer agreement between pathologists. 
	}   
	}
	\vspace{-4mm}
	\label{fig:quantEval}
\end{figure*}

\noindent \textbf{Evaluation of AI-FFPE's efficiency in generating FFPE-style images from FS images}
\label{ssec:evalmetric}\\
To further corroborate AI-FFPE's image translation efficiency we designed a Visual Turing test performed by board certified pathologists. The aim of the test was to determine if expert eyes can distinguish AI-FFPE generated FFPE-style images from real FFPE images. In this Visual Turing test, 20 pathologists were shown AI-FFPE generated and real FFPE tissue section patches at a random sequence and were asked to decide if the images were synthetically transformed from frozen images or they were real FFPEs. Each pathologist was shown the same survey which is comprised of 25 randomly selected FFPE patches from test dataset and 25 AI-FFPE generated patches. The images were shuffled using random number generator functions. The results are summarized in \textbf{\Cref{fig:quantEval}b} and shown as equally weighted harmonic average of precision and recall values, $F_1$ score. While 55.6\% of brain and 56\% of lung AI-FFPE generated patches were classified as real FFPE, 49\% of brain and 43.8\% of lung real FFPE images were classified as real FFPE by pathologist, marking the outstanding efficiency of our model in translating the images to the FFPE domain. We also evaluated if pathologist agreed on which images should be classified into each group using Fleiss' kappa metric \cite{Fleis1971}, a measurement between 0 and 1, 1 indicating full inter-observant consensus. Fleiss' kappa values of 0.024 for lung and 0.058 for brain very poor inter-observer agreement in both surveys, strongly suggesting that images were randomly allocated to each group by pathologists, further supporting the results that demonstrates expert eyes were unable to distinguish AI-FFPE generated images from real FFPE images.

\begin{figure*}[!b]
\includegraphics[width=\textwidth]{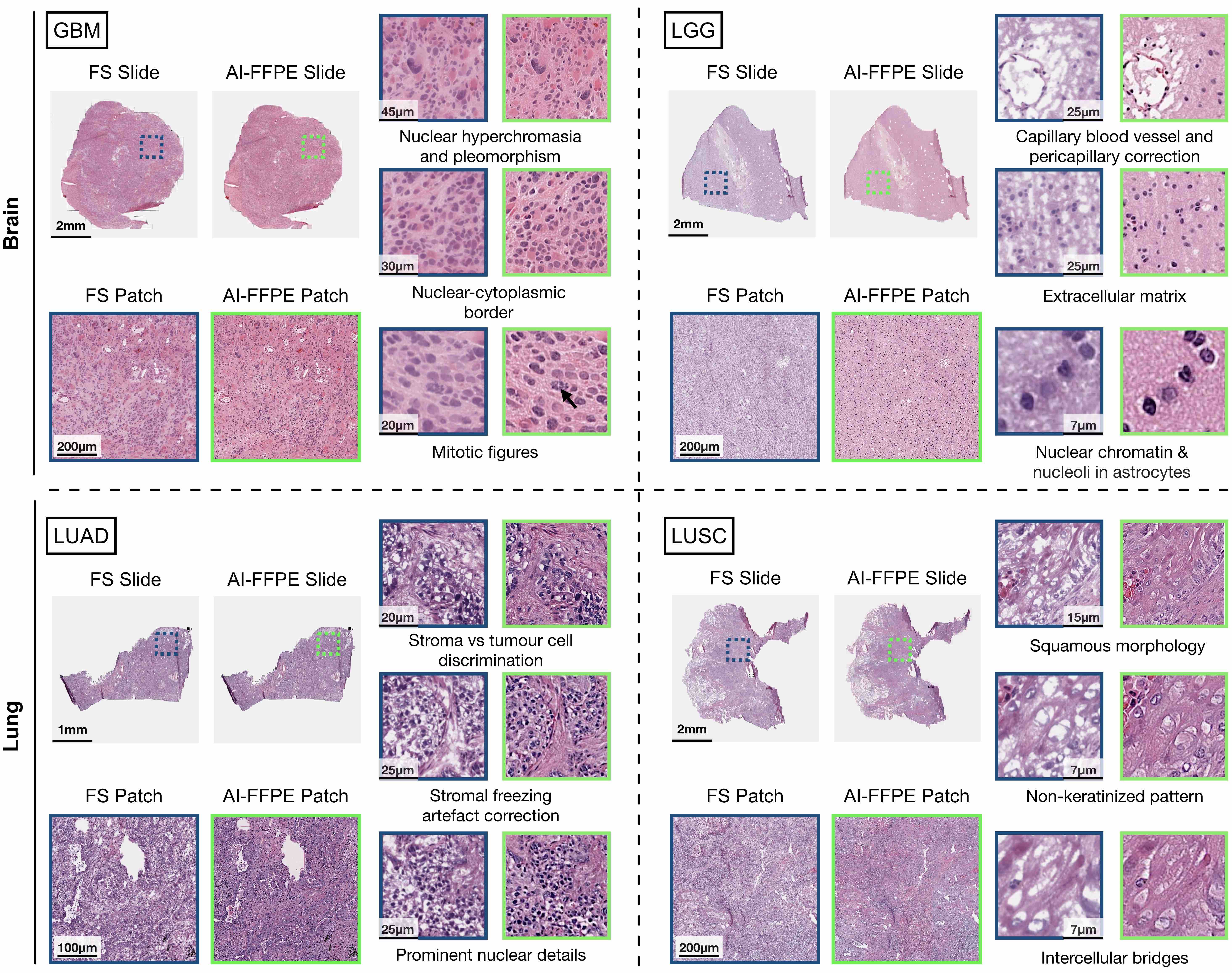}
	\caption{\textbf{ AI-FFPE enhancement of tumor-specific diagnostic patterns.} Comparison of FS vs. AI-FFPE images demonstrating how AI-FFPE reveals and augments diagnostic characteristics in Glioblastoma Multiforme (GBM), Low-grade Gliomas (LGG), Lung Adenocarcinoma (LUAD), and Lung Squamous Cell Carcinoma (LUSC) slide images. AI-FFPE's FS artefact correction methods such as image sharpening, rectification of empty spaces helps clearer identification and evaluation of nuclear borders, diffuse growth pattern, development of new capillaries (angiogenesis), necrosis, and mitotic figures allowing more precise grading and easier diagnosis of atypia for brain tumors. Similarly, AI-FFPE transformation of lung patches makes squamous and adenocarcinoma-specific morphology more visible by correcting stromal freezing artefacts, producing more prominent nuclear details, stroma/tumor distinctions for the former; highlighting epithelial architecture and cellular characteristics for the latter.}
	\vspace{-4mm}
	\label{fig:stitch}
\label{Fig:StitchResults}
\end{figure*}

We further examined if AI-based FS-to-FFPE image transformation leads to improvement of diagnostically valuable visual patterns, particularly of those that deteriorate due to FS. \textbf{\Cref{fig:stitch}} compiles examples of improved visual patterns of importance that are specific to each of the four cancer types that we examined in this study. Increased nuclear colour intensity (hyperchromasia), diverse nuclear morphology (pleomorphism), substantially increased mitotic rate manifested with the presence of mitotic figures are the differentiating histological features of GBM, the highest grade of gliomas. Lower grade gliomas, however, has to be differentiated from gliosis which is a benign condition. Nevertheless, in many cases, differential diagnosis based on H\&E staining alone could be difficult and requires further molecular characterisation which is a time-consuming process to inform intra-operative decision-making. In other cases of LGG, presence of certain neoplastic patterns, such as nuclear atypia, mitotic figures, microvascular proliferation can inform diagnosis with H\&E slides alone. AI-FFPE makes such patterns in FS images more noticeable by improving appearance of the smallest vessels (capillaries) and areas surrounding these vessels (pericapillary area), ECM and structures in the nuclei (chromatin and nucleoli). In lung, AI-FFPE also improves the clarity of diagnostic patterns for adeno- and squamous carcinomas. For adenocarcinomas, rectification of freezing artefacts in tumour stroma, enhancing nuclear details of the tumour cells and overall improvement in distinction of stroma and tumour cells making features of diagnostic significance more visible. On the other hand, for squamous carcinomas, AI-FFPE's improvement of epithelial features, such as, squamous appearance, non-keratinized pattern and presence of intercellular bridges makes squamous cancer-specific features more prominent and easily recognisable. In summary, these improvements show AI-FFPE's ability to repair and enhance cancer-type specific diagnostic patterns.

\noindent \textbf{Adapting networks trained on FFPE to Frozen Section.}

Finally, we assessed if the results we presented above would render an increased diagnostic performance by comparing classification performance of our recently published CLAM algorithm\cite{lu2020data} with the inputs of AI-FFPE pre-processed and regular FS WSIs. For classification of non-small cell lung cancers into adenocarcinoma or squamous cell carcinoma subtypes, CLAM achieved significantly higher test AUC of 0.952 on AI-FFPE WSIs, compared to AUC of 0.9061 on FS WSIs. For the subtyping of gliomas as second (LGG) and fourth (GBM) grade tumours, the classification model has benefited from AI-FFPE's image improvements, achieving 0.9837 AUC score for AI-FFPE WSIs, significantly higher than 0.9122 AUC score for FS WSIs (Supplementary Fig. 4). 

\begin{figure*}[!b]
\includegraphics[width=1\textwidth]{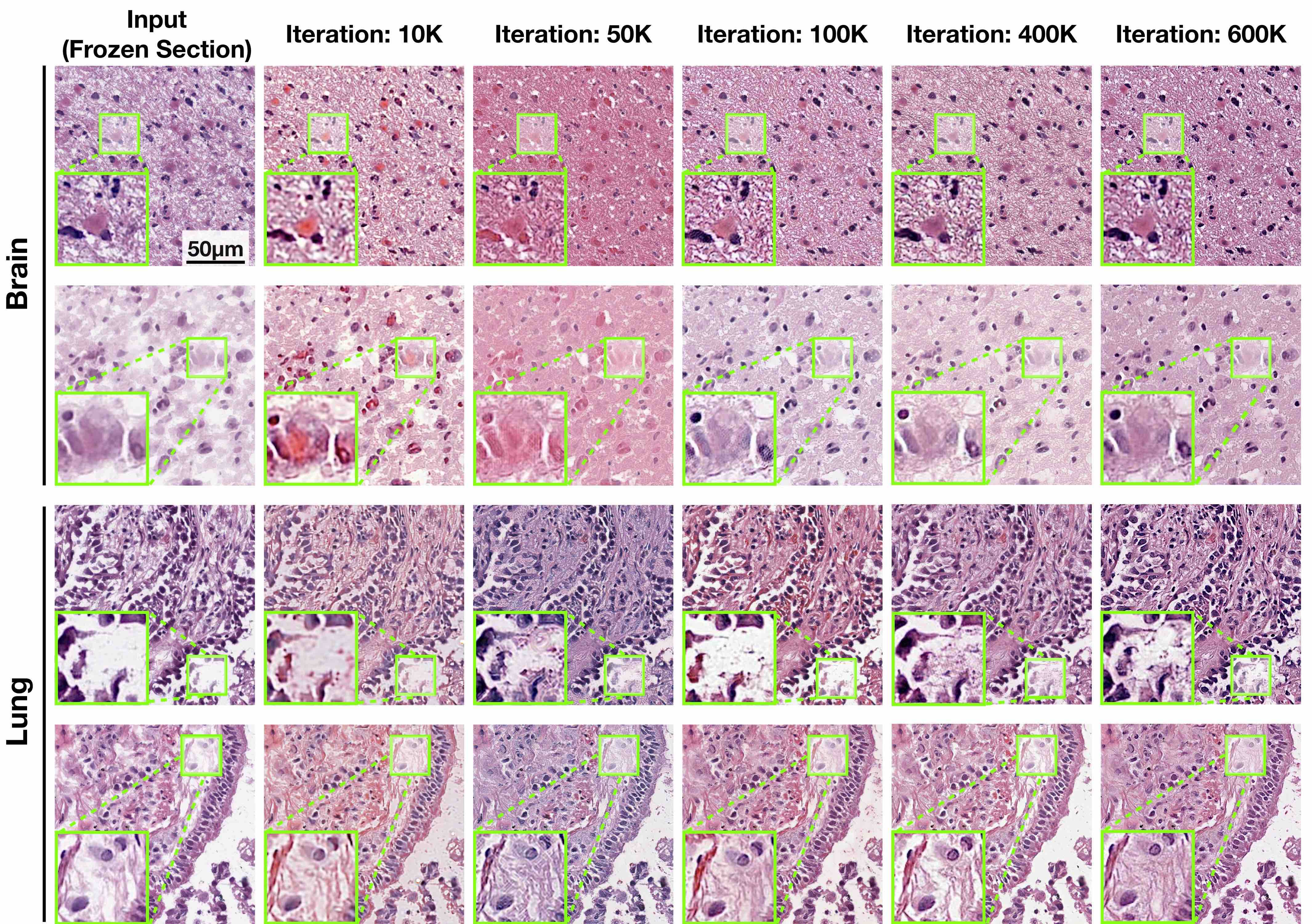}
\caption{\textbf{Improvement of output image quality throughout the network training.} Network output images for the brain and lung tissue section at different stages of the learning process, that is, after 10k, 50k, 100k, 200k, 400k and 600k In brain sections, the visibility of astrocytic glial neoplastic and stromal cell nuclei, as well as the fibrillar structures
improve through iterations. Even though the visual enhancement in the lung tissue sample is highly challenging due to alveolar architecture, significant restoration of the connective tissue is observed as the training progress. At the beginning of training, the diagnostically misguiding regions such as artificial presence of bleeding and blurred nuclear boundary were frequently observed in AI-FFPE patches. However, these issues have been resolved at the end of five epochs.  
} 
\label{fig:iteration}
\end{figure*}

\begin{figure*}[!h]
\includegraphics[width=\textwidth]{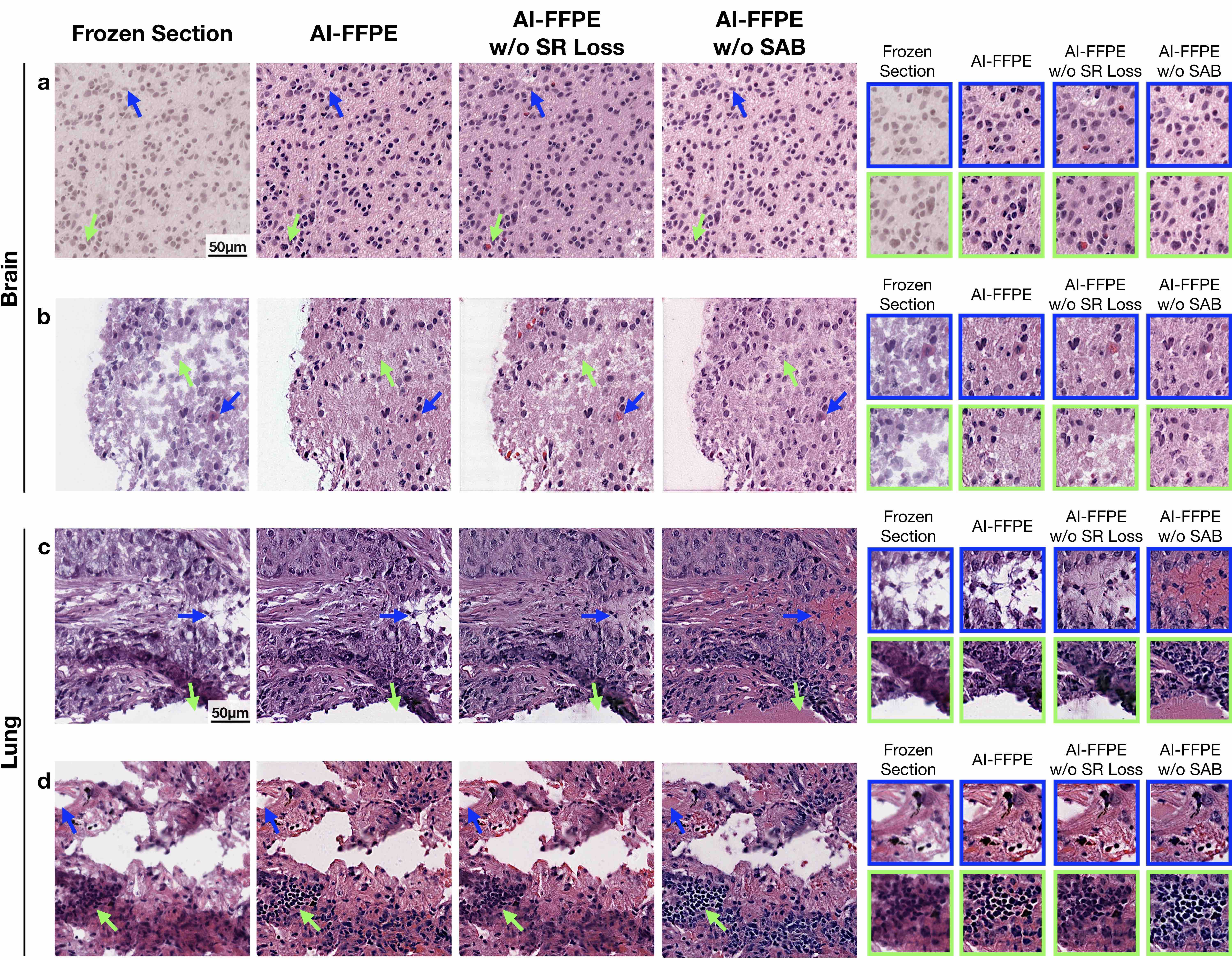}
\centering
	\caption{\textbf{SAB and self-regularization (SR) loss function effect on the model performance.} The implementation of SR-loss to the objective function prevents the network from adding clinically misguiding components such as erythrocytes (\textbf{\color{green}{a}}, \textbf{\color{blue}{b}}) which might give a false-impression of haemorrhage, enhances the appearance of nuclear borders (\textbf{\color{green}{d}})  and improve overall staining quality (\textbf{\color{blue}{a}}). SR-loss implementation also repairs blank areas, however, without including textural details such as the presence of fibrils and lacking sufficient level of selectivity.  With the integration of SAB module, the model gains the ability to detect the regions (\textbf{\color{green}{b}}, \textbf{\color{blue}{c}}) where extracellular matrix (ECM) modifications should be avoided. When SR-loss and SAB are combined, the images show not only cumulative improvement but also synergistic effect such as, adding fibrils and other textural components to ECM and filling the areas that are lost by frozen sectioning (\textbf{\color{green}{c}}). Rarely, SR-loss and SAB antagonises each other in image improvement, for example SR alone is sometimes more efficient in enhancing nuclear appearance compared to SR loss-SAB combination (\textbf{\color{green}{c}}, inside the picture frame). AI-FFPE also make lymphocytes more prominent, hence, easier to distinguish from the tumor cells (\textbf{\color{green}{d}}). SAB module also prevents the model from filling the vessels' lumen with ECM (\textbf{\color{blue}{d}}).}
	\vspace{-4mm}
	\label{fig:ablation}
\end{figure*}

\section*{Discussion}
\label{sec:discussion}
\vspace{-2mm}
Computational histopathology is a promising field that has been increasingly offering novel tools to improve and evaluate microscopic tissue images and growing number of studies focusing on histopathological applications of AI have been published so far. These studies range from developing new tools to assist pathologists by automatised counting elements of interest on tissue slides to computational algorithms that guide the pathologist to diagnostically relevant regions; attempts to replace chemical staining of tissues with virtual staining to tumour sub-typing. However, improvement of frozen section image quality which is inherently susceptible to deterioration because of the the sample preparatory techniques has remained as an unexplored territory. In this study, we developed an AI algorithm called AI-FFPE, that efficiently resolves many frozen image quality related issues in frozen-sectioned brain and lung tissues. Frozen sectioning introduces artefacts with diverse characteristics. Some artefacts obscure cellular and ECM details, a few others (folding, chattering and thickness variation) produces uneven cellular density; while freezing artefacts introduce artificial blank spaces into the ECM and the cells, staining artefacts compromise colouring quality. Our method's efficiency in correcting aforementioned artefacts of different nature demonstrates its versatility which is also reflected in its ability to highlight patterns of diagnostic importance in different types of tumours and tissues. For example, AI-FFPE's transformation enhanced such diagnostic patterns for lung adeno- and squamous carcinomas, both are epithelial in origin but exhibit distinct patterns of diagnostic importance. While these patterns cover both stroma and epithelium in the lung adenocarcinomas, they are more concentrated in the epithelium itself for squamous lung cancers. Despite the distinct architecture and embryonic origins of brain gliomas from the lung carcinomas, AI-FPPE also improves the appearance of diagnostically significant patterns in brain glioma WSIs. These results suggest great prospects for application of AI-FFPE to WSIs of other tumours and organs, and WSIs with other artefactual alterations than we assessed here. Further studies involving other tumours of brain and lung, other organs and artefact types are required to confirm malleability of our method to these circumstances. 

Another strength of our study is that we did not examine our method only in isolation but we also performed comparative studies with generic image translation models, such as CUT, FastCUT and CycleGAN. Our model showed significant superiority in FID scores and output images of slide patches that harbour different types of artefacts. As shown in the relevant figures, integration of SR-loss and SAB to final AI-FFPE version greatly contributed to these remarkable results.    

Finally and crucially, our results not only establish outstanding improvements in the images but also demonstrate that these improvements culminate in increased diagnostic accuracy by showing that AI-FFPE transformed images significantly increase CLAM's cancer sub-typing performance. In the future, prospective clinical studies can validate AI-FFPE's contribution to intra-operative diagnostic accuracy of FS samples and surgical decision-making in real hospital settings.

\end{spacing}

%
%
\vspace{-5mm}
 \section*{\large{Online Methods}}
\begin{spacing}{1}
\vspace{-4mm}

To deal with the problems related to FS examinations, we propose an AI-FFPE method that transforms pathology sections while retaining clinically relevant information without inserting any semantically misleading structures. In traditional unpaired image-to-image translation methods, cycle-consistency dictates the similarity between the images from the target domain and the reconstructed images generated by inverse mapping. The consistency is enforced under different aspects, e.g.,image to latent \cite{lee2018diverse,zhu2018multimodal}, latent to image \cite{MUNIT2018} or two-image domains \cite{kim2017learning}. In the typical state-of-the-art applications, adversarial loss \cite{Goodfellow2014} pushes the change in target appearance while the content is preserved by the cycle consistency loss. At a higher order, the model examines the patches from source and target domains and penalizes the discrepancy on the diagnostically relevant regions between the frozen and synthesized patches to improve the quality of vision by preserving the content. This paves the way for optimal stain transfer between domains, as it can be easily isolated from content-related features. However, the bijectivity assumption behind cycle consistency restricts the model especially in cases where a large number of uncommon features exist across domains. DistanceGAN \cite{DistGAN2017}, TraVeLGAN \cite{amodio2019travelgan}, and GcGAN \cite{GcGAN2019} propose one way translation to overcome the circularity-based constraints.  As an alternative approach, UNIT \cite{Liu2017} and MUNIT \cite{MUNIT2018} propose to learn a common content latent space by decomposing images into domain-invariant content representation and the domain-specific identities with respect to the style-code. However, defining an objective function based only on a high-level signal that works for pixel-wise reconstruction leads to high computational complexity as well as blurry output images.  Although Park et al. \cite{park2020contrastive} proposes a patch-based approach to avoid these burdens, it has a major bottleneck for applications in clinical pathology in cases where a balance between rectification of artefacts without changing the cellular formation is needed. 

\noindent\textbf{Dataset Description} \\
  The Cancer Genome Atlas (TCGA) open-source database was used to train and test AI-FFPE algorithm. For brain, TCGA-GBM (highest grade brain tumours) and TCGA-LGG (low grade brain tumours) datasets, made of tumours that belongs to the most common cancer -gliomas- of brain and representing two distinct histopathological, biological and clinical patterns were used. For lung, TCGA-LUSC and TCGA-LUAD projects which are composed of WSIs from two most common but histologically distinct lung cancer types were utilised. The subset of these projects been used.  We utilised a subset of dataset from these projects. Our subset consisted of 97,271 (X GBM, Y LLG) frozen and 110,087 (X GBM, Y LLG) FFPE patches from 590 (X GBM, Y LLG) patients for brain; 135,785 (X LUSC, Y LUAD) frozen and 71,311 (X LUSC, Y LUAD) FFPE patches from 650 (X LUSC, Y LUAD) patients for lung.

\vspace{-4mm}
\noindent\textbf{WSI Processing} \\
The regions each biopsy slide where tissue was present were first segmented and stored as object contours. Then, we extracted 512 $ \times $ 512 - sized image patches without spatial overlapping from 20 $ \times $ magnification using the segmentation contours.\\
\noindent \textbf{\textit{Segmentation and Patching of WSI.}} 
The biopsy tissue in each WSI were segmented using the CLAM WSI-analysis toolbox \cite{lu2020data}. A binary tissue mask denoting the tissue and non-tissue regions were computed for each  downsampled input image in HSV color space by thresholding the saturation channel median blurring. The estimated contours of the denoted tissue and the cavities of tissue were then filtered depending on their area to generate the final segmentation mask. The model was trained on 20$\times$ magnifications which was segmented into 512 $\times$ 512 patches without overlap.\\
\noindent\textbf{Method Architecture} \\
\noindent \textbf{\textit{Hyperparameters and Training Details.}} 
After randomly sampling slides, we have trained a GAN under the supervision of adversarial, contrastive and self-regularization loss using a mini-batch size of one patch. To avoid adding clinically irrelevant information to the images, we enforced the network to create target domain synthetic FFPE-like images having a closer content to frozen samples using a self-regularization loss with the aim of preserving the spatial orientation of the nuclei and other diagnostically relevant features, whereas intensity and ratio of the integrated self-regularization functionality is adjustable by a weight hyperparameter. Unlike the typical GAN network architecture that consists of ResNet-based generator and PatchGAN discriminator with the Least Square GAN loss of contrastive unpaired loss translational learning (CUT) models \cite{park2020contrastive}, our architecture employs a generator with artefact-aware attention block and patch-based self-aware contrastive loss. Since our method simplifies the training procedure by operating just in one direction, that is from source domain to target, training time drastically decreases compared to traditional cycle consistency-based unpaired image-to-image translation methods in literature.
Standard adversarial loss is the first component of our hybrid loss function:
\begin{equation}
\mathcal{L}_{GAN}(G, D, \mathbb{X},\mathbf{Y}) = \mathbb{E}_{y \sim \mathbf{Y}} \log D(y) + \mathbb{E}_{x\sim  \mathbf{X}} \log(1-D(G(x))),
\label{eq:gan}
\end{equation}
where the discriminator $\mathbf{D(y)}$ attempts to recognize if the patch is a synthetically generated AI-FFPE sample by $\mathbf{G(x)}$ or a real FFPE sample, $\mathbf{y}$. The adversarial loss function dictates to learn and eliminate the style differences, which triggers the use of a noise contrastive estimation function that guarantees the content preservation in patch level \cite{NCE2010}. Therein, we employ a patchwise noise contrastive loss to find the resemblance between input FS and output image AI-FFPE patches, by taking a query patch from a generated image in the FFPE target domain and match it with a corresponding frozen sample image patch at the same location with an expectation that they will form a positive pair, whereas other patches that are dissimilar will form negative pairs. 
One positive, and $M$ negative examples are being mapped to $L$-dimensional real vector space $v, v^ + \in \mathbb{R}^{L}$ and $v^{-}\in \mathbb{R}^{M\times L}$, individually. $v^{-}_k \in \mathbb{R}^{L}$ signifies the k-th negatives in $M$ samples. We normalize the vectors onto unit spheres to stop the space from collapsing or even growing. Problem is set as $(N+1)$-ways classification, where scaling the distance between the query and the examples by a temperature $\nu=0.07$ also passing as logits~\cite{wu2018unsupervised}.  The probability of positive examples selected over the negative ones is formulated as a cross-entropy loss and calculated as:
\begin{equation}
\mathcal{L}(v, v^+, v^-) = -\log{ \Bigg[
\frac{\exp ( v . \ v^+/ \nu)}{\exp (v . v^+/ \nu) + \sum_{n=1}^N \exp (v . v^-_n/\nu)} \Bigg] }.
\label{eq:nce_basics}
\end{equation}
As per defined probability, $\mathcal{L}(v, v^+, v^-)$, the patches from the frozen section tissue boundary are expected to be more closely associated with the synthesized FFPE section's boundary than the patches from other regions of the slide. Once L-layers of interest are selected from the generator, feature maps coming from generator are given as input to the Multi-Layer Perceptron (MLP) as introduced in SimCLR\cite{chen2020}.  Similarly, synthesized images are encoded with these two networks as $\{ \hat{z} \}_L=\{H_l(G_{enc}^l(G(x)))\}_L$ where $l \in \{1, ..., L\}$ features, and patch noise contrastive estimation loss is defined based on the final features $\{z_l\}_L's$:
\begin{equation}
\mathcal{L}_{PatchNCE}(G, \mathbf{F}, \mathbf{X}) = \mathbb{E}_{x \sim \mathbf{X}} \sum_{l=1}^L \sum_{s=1}^{S_l} \mathcal{L}(\hat{z}_l^s, z_l^s, z_l^{S\setminus s}).
\label{eq:NCEloss}
\end{equation}
To prevent the network from adding any clinically misleading information, we penalize the significant deviations from real input images by a self-regularization pixel-wise $L_1$ loss function: 
\begin{equation}
\mathcal{L}_{sReg}(G,X) = ||\mathbf{X}-G(\mathbf{X})||_1
\end{equation}
 For each patch, the final objective is a weighted sum of these loss functions in total:
\begin{equation}
\mathcal{L}_{AI-FFPE} = \mathcal{L}_{GAN}(G,D,\mathbf{X}, \mathbf{Y}) + \lambda_{sReg} \mathcal{L}_{sReg}(G,X) + \lambda_{\mathbf{X}}\mathcal{L}_{patchNCE}(G,\mathbf{F}, \mathbf{X}) + \lambda_{\mathbf{Y}}\mathcal{L}_{patchNCE}(G, \mathbf{F}, \mathbf{Y}).
\end{equation}
That way, we aim to achieve diagnostically more informative and interpretable images after translation by the network under the $\mathcal{L}_{AI-FFPE}$ supervision. Our model was trained using the Adam optimizer\cite{kingma2017adam} with an initial learning rate of $0.002$ as well as the momentum parameters $\beta_1=0.5$ and $\beta_2=0.999$ for $5$ epochs and the outputs throughout the iterations are given in \textbf{\Cref{fig:iteration}}. We use a batch size of $1$, instance normalization \cite{Ulyanov2016}, and Xavier weight initialization \cite{article}. Res-Net with $9$ residual blocks \cite{DBLP:journals/corr/HeZRS15} is chosen as a generator, PatchGAN as discriminator \cite{Isola2016}, and Least Square GAN loss \cite{8237566}. Accordingly, we set $\lambda_{NCE}$ as $1$ for identity loss with temperature $(\tau)$ $0.08$, and enqueued $512$ patches for each image for each iteration. 


\noindent \textbf{\textit{Spatial Attention Block (SAB).}}
The SAB mechanism can be seen as a non-local convolution operation extracting the relative weights of all positions on the feature maps for any given input $\mathbf{X} \in \mathcal{R}^{N\times64\times H \times W}$:
\begin{equation}
\mathbf{Z}=\mathit{f}(\mathbf{X}, \mathbf{X}^\top)\mathit{g}(\mathbf{X}),
\end{equation}
where $\mathit{f}$ stands for the pixel-wise relations of input $\mathbf{X}$. After convolving the block inputs, we employ the dot product operation on the final tensor of max-pooled $\phi$ layers, which are activated by Rectified Linear Unit (ReLU) function,$\sigma_{relu}$ :
\begin{equation}
    \mathbf{P} = \psi(\sigma_{relu}(\theta(\mathbf{X})\phi(\mathbf{X})^\top)).
\end{equation}
The dot product, $\theta(\mathbf{X})\varphi(\mathbf{X})^\top$, provides input covariance measurement, which can be interpreted as a degree of inclination between two feature maps at different channels. We activate the $\psi$ convolution operation in $\mathit{softmax}$ function and perform a matrix multiplication between the $g$ and the output of $\mathit{softmax}$. Then, the result of $\phi$ multiplication is convolved and upsampled to produce the $\mathbf{S}$. Finally, an element-wise sum operation between attention map $\mathbf{S}$ and the input $\mathbf{X}$  generates the output $\mathbf{E} \in \mathbb{R}^{N\times64\times H\times W}$:
\begin{equation}
    \mathbf{S} = \phi (\sigma_{softmax}(\mathbf{P})g(\mathbf{X})),
\end{equation}
\begin{equation}
    \mathbf{F} = \mathbf{S} + \mathbf{X}, 
\end{equation}
where $\sigma_{softmax}$ denotes $\mathit{softmax}$ function. Short connection between the input $\mathbf{X}$ and the output $\mathbf{F}$ finalizes the block by strengthening the residual signals. The detailed flow diagram of block operations of SAB module is given in \textbf{\Cref{fig:summary} c}.  


\noindent \textbf{\textit{Patch-level Evaluations.}}
For the quantitative assessment of the GAN performance at patch-level, the Fréchet Inception Distance (FID) \cite{heusel2018gans} is utilized as a metric that is most compatible with human perception. The multidimensional Gaussian distribution of the real FFPE images and the generated AI-FFPE images in a deep network space as well as the difference of mean and standard deviations of these two distributions are computed. As the generated images start becoming more realistic throughout the iterations, their statistics resemble to the real FFPE images from target domain and FID score decreases gradually.
The FID, $\mathcal{I}_{FID}$, can be formulated as: 
\begin{equation}
\mathcal{I}_{FID} = ||\vec{\mu_1}-\vec{\mu_2}||_2^2 + Tr(C_1+C_2-2*\sqrt{(C_1*C_2}))
    \label{eq:FID}
\end{equation}
where $\mu_1$ and $\mu_2$ refer to the feature-wise mean of the real and generated images, $C_1$ and $C_2$  are co-variance matrices for the real and generated feature vector, $Tr$ stands for trace function, e.g. the sum of element along the main diagonal of the square matrix. 

\noindent \textbf{\textit{Slide-level Evaluations.}}
After processing frozen patches individually, we are stitching the AI-FFPE patches back to reconstruct the whole slide image. The model is trained at the patch-level, which allows the training dataset size to be scaled-up and also makes the process memory efficient. Inevitably, the slide-level examinations that are performed by pathologists introduce certain levels of noise to the evaluation due to the inter-observer differences. Nonetheless, we showed that it is possible to train a robust unpaired model to enhance the diagnostic vision quality. This approach is highly compatible with the clinical workflow, where diagnosis are made per slide and not for a particular region of slide separately. By mitigating the burden of making diagnosis on inferior slides or time delay rooted from the need for a second sectioning, the slide evaluation time decrease with the increase of interpretability in clinical applications.

\noindent\textbf{Computational Hardware and Software.} \\
WSIs were processed on Intel Xeon multi-core CPUs (Central Processing Units) and 2 NVIDIA 2080 Ti GPUs (Graphics Processing Units) using the  publicly available CLAM\cite{lu2020data} whole slide processing pipeline implemented in Python (version 3.7.5). Deep learning models were trained on NVIDIA GeForce RTX 3090 GPUs using Pytorch (version 1.7.0). 


\vspace{-6mm}
\section*{Data Availability}
\vspace{-6mm}
All reasonable requests for refined slides will be evaluated by the authors to decide if the demand is subject to any confidentiality or intellectual property obligations. All requests for data that can be shared will be processed through formal channels, in agreement with departmental and institutional guidelines and will require a material transfer contract. 
\vspace{-4mm}

\section*{Code Availability}
\vspace{-6mm}
All code of this paper that is implemented using PyTorch in Python are available at  \url{https://github.com/DeepMIALab/AI-FFPE} to reproduce the experiment results.

\vspace{-6mm}

\section*{Author Contributions}
\vspace{-4mm}
 M.T., F.M. and K.B.O conceived the study and designed the experiments. K.B.O and G.I.G performed the experimental analysis. B.D., M.K., M.Y.L., and T.Y.C. curated training and test dataset. K.B.O, S.C., K.B., D.D., G.S., M.T, and F.M. analyzed the results. K.B.O, S.C., U.P.H., D.F.K.W., M.T and F.M. prepared the manuscript. M.T. and  F.M. supervised the research.

\vspace{-4mm}
\section*{Acknowledgements}
\vspace{-6mm}
Mehmet Turan, Kutsev~Bengisu~Ozyoruk, 
 Guliz Irem Gokceler, and Mohamad Kassab are especially grateful to the Scientific and Technological Research Council of Turkey (TUBITAK) for International Fellowship for Outstanding Researchers.

\vspace{-6mm}
\section*{Competing Interests}
\vspace{-6mm}
The authors declare that they have no competing financial interests.
\vspace{-6mm}

\end{spacing}


\section*{References} 
\vspace{2mm}

\begin{spacing}{0.9}
\bibliographystyle{naturemag}

\end{spacing}


\newpage
\begin{extfigure*}
\includegraphics[width=\textwidth]{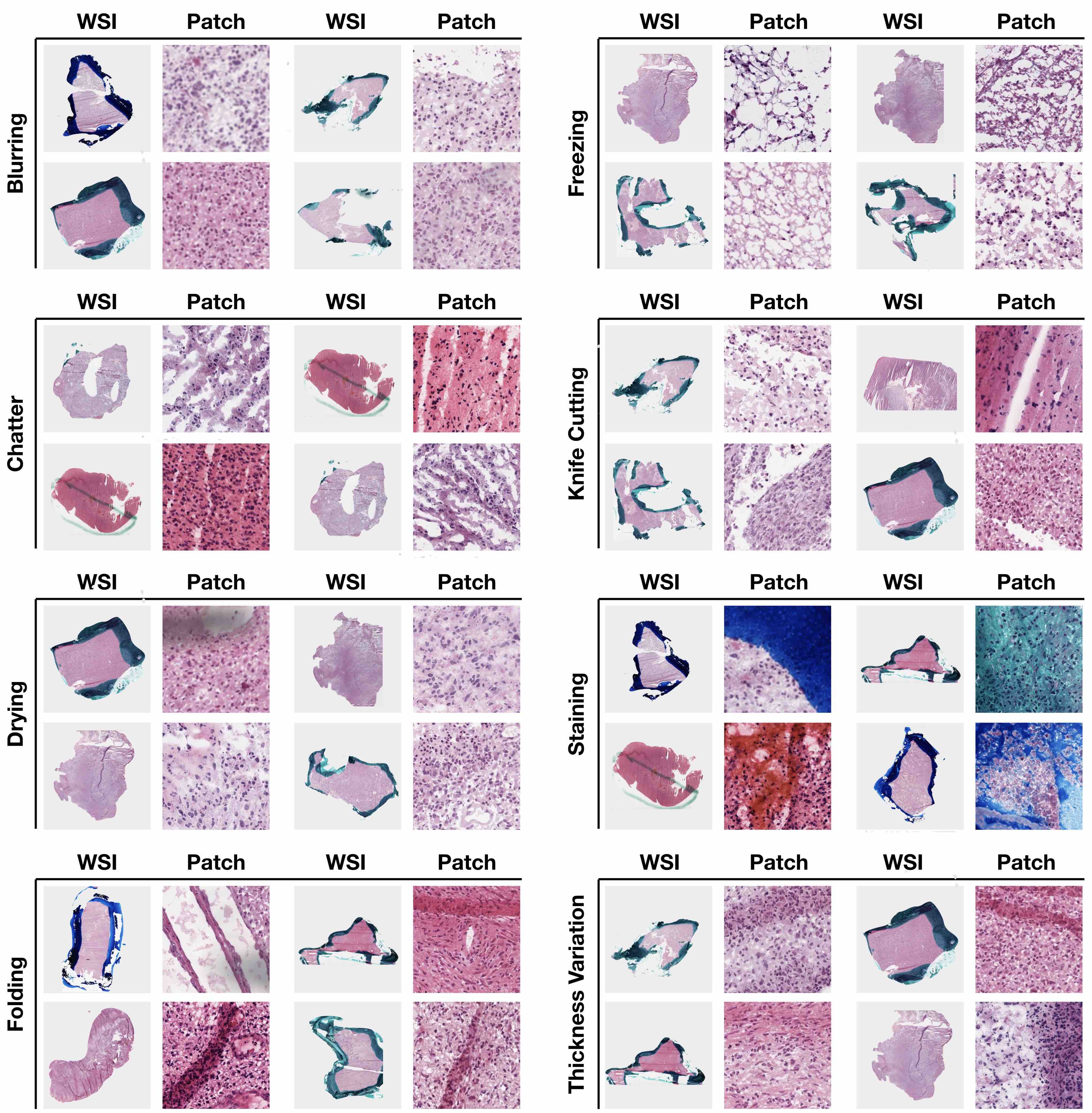}
\caption{\textbf{Frozen Artefacts.} Artefacts of brain tissue preparations, belonging to eight different classes, four examples from each class, are presented here. Chatter and knife cutting are the forms of mechanical damage produced during the tissue cutting. Folding is caused by the erroneous mounting of the tissue sections on the slides. Over-staining is caused by leaving the sections in the staining solutions longer than needed, insufficient washing following the staining or use of overly diluted alcohol solutions. When slightly more than prescribed time is taken between removing the slide and applying the coverslip, the section starts to dry causing drying artefacts. Freezing artefacts which occur due to fast/slow or under/over congealment are unique to the frozen section procedure while many other artefacts are encountered in both paraffin and frozen sections. }
\label{fig:artefacts}
\end{extfigure*}

\newpage
\begin{extfigure*}
\includegraphics[width=\textwidth]{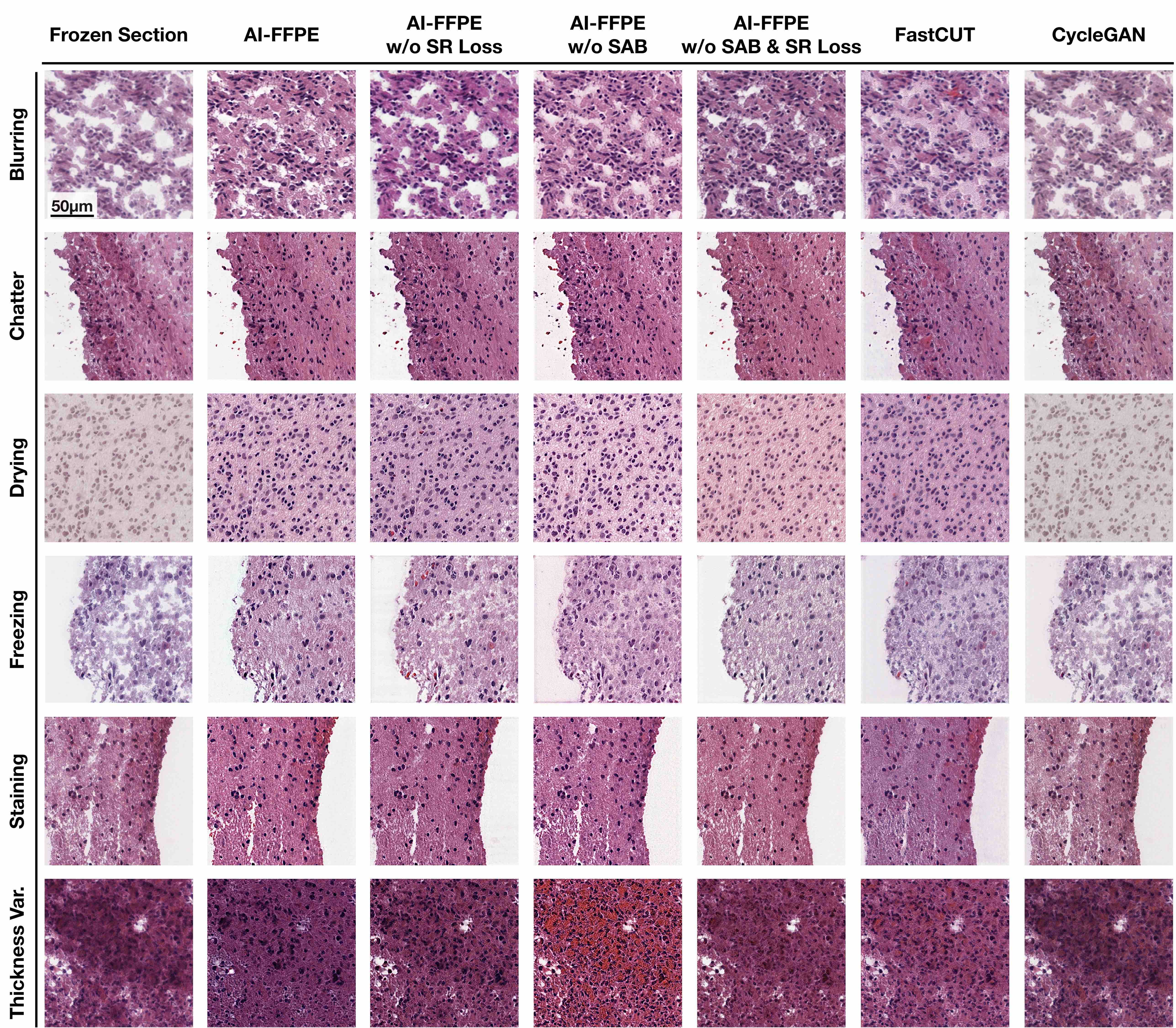}
\caption{\textbf{Comparison of all bench-marked methods' improvement of various artefacts in brain tissue sections} AI-FFPE (final version) are compared to several unsupervised image-to-image translation methods such as CyleGAN, FastCUT, AI-FFPE without spatial attention block integration (AI-FFPE w/o SAB),  AI-FFPE without self-regularization loss integration (AI-FFPE w/o SR loss),  AI-FFPE without both self-regularization loss and SAB (AI-FFPE w/o SAB $\&$ SR loss)}
\label{fig:all_models_brain}
\end{extfigure*}

\newpage
\begin{extfigure*}
\includegraphics[width=\textwidth]{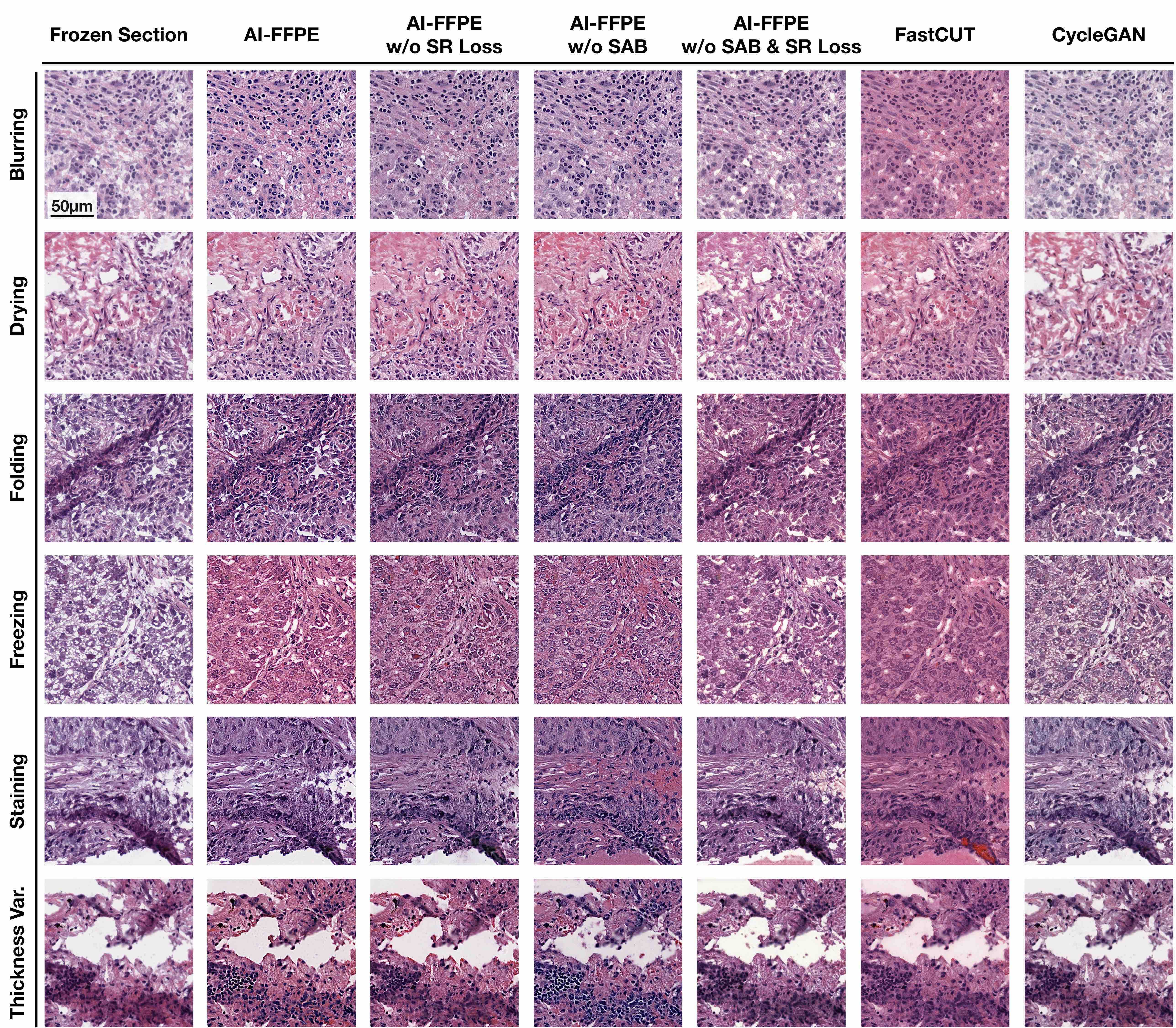}
\caption{\textbf{Comparison of all bench-marked methods' improvement of various artefacts in brain tissue sections.} Under the constrain of cycle consistency loss, CycleGAN, in most of the cases, does not implement any changes on the input FS images. Also, FASTCUT's contrastive learning is useful to maximize shared content-related features in between input and synthesized image patches, however, lacks an essential quality required for the improvement of lung WSIs and that is to determine the tissue edges, such as boundaries of vessels or airways, beyond which has to remain untouched.}
\label{fig:all_models_lung}
\end{extfigure*}
\end{document}